\documentclass[a4paper,11pt]{article}
\usepackage{pos}
\usepackage{amsmath}

\usepackage{amssymb}
\usepackage{anyfontsize}
\usepackage{subfigure}

\DeclareMathAlphabet{\mathcal}{OMS}{cmsy}{m}{n}

\graphicspath{{./}}

\renewcommand{\l}{\left}
\renewcommand{\r}{\right}
\newcommand{\g}[1]{\gamma_{#1}} 
\newcommand{\tr}{\mathrm{tr}}

\newcommand{\chiral}[1]{\mathring{#1}} 

\newcommand{\gev}{\,\mathrm{GeV}}

\newcommand{\mev}{\,\mathrm{MeV}}
\newcommand{\fm}{\,\mathrm{fm}}

\newcommand{\SU}[1]{\mathrm{SU}\l(#1\r)}

\newcommand{\phys}{\mathrm{phys}}

\newcommand{\tins}{t_\mathrm{ins}}

\newcommand{\tsep}{t_\mathrm{sep}}

\newcommand{\vp}{\mathbf{p}}
\newcommand{\vpi}{\mathbf{p}_i}
\newcommand{\vq}{\mathbf{q}}
\newcommand{\vpf}{\mathbf{p}_f}
\newcommand{\vxi}{\mathbf{x}_i}
\newcommand{\vxop}{\mathbf{x}_{op}}
\newcommand{\vxf}{\mathbf{x}_f}

\title{The pion scalar form factor with $N_f=2+1$ Wilson fermions}

\author*[a]{Konstantin~Ottnad}
\author[a]{Georg~von~Hippel}
\affiliation[a]{PRISMA\textsuperscript{+} Cluster of Excellence and Institut f\"ur Kernphysik, Johannes Gutenberg-Universität Mainz, \\
Johann-Joachim-Becher-Weg 45, 55099 Mainz, Germany}

\emailAdd{kottnad@uni-mainz.de}

\abstract{We report preliminary results from an analysis of the pion scalar form factor computed on a set of the $\mathrm{tr}[M]=\mathrm{const}$ CLS gauge ensembles with $N_f=2+1$ Wilson Clover-improved sea quarks. The calculations are carried out for light quarks masses corresponding to $M_\pi\approx 0.130\mathrm{MeV} \ldots 350\mathrm{MeV}$, four values of the lattice spacing $a\approx0.049\fm\ldots0.086\fm$ and a large range of physical volumes. A fine-grained momentum resolution is achieved by allowing for non-vanishing sink momenta and by including two particularly large and fine boxes close to physical quark masses (i.e. $T\times L^3 =192\times 96^3$, $M_\pi \le 172\mathrm{MeV}$, $a \le 0.064\fm$). The pertinent quark-disconnected contributions have been computed to high precision using a scheme combining 1.) the one-end trick on stochastic volume sources for the computation of differences between two quark flavors with 2.) the hopping parameter expansion and hierarchical probing to evaluate the loops for the heaviest, single quark flavor. \\ \vspace{1em} \flushright{MITP-24-003}}

\FullConference{The 40th International Symposium on Lattice Field Theory (Lattice 2023)\\
July 31st - August 4th, 2023\\
Fermi National Accelerator Laboratory\\}

\begin{document}
\maketitle

\section{Introduction}
The scalar form factor of the pion
\begin{equation}
 F_S^{\pi,l}(Q^2) = \langle \pi(p_f) | m_d \bar{d}d+m_u\bar{u}u| \pi(p_i)) \rangle \,,
 \label{eq:F_S_pi_l}
\end{equation}
is not directly accessible to experiment. Still, the corresponding radius
\begin{equation}
 \langle r_S^2 \rangle_\pi^l = \frac{-6}{F_S^{\pi,l}(0)}\l.\frac{dF_S^{\pi,l}(Q^2)}{dQ^2}\r|_{Q^2=0} \,,
 \label{eq:rsqr_pi_l}
\end{equation}
is relevant to the low-energy regime of QCD as it is related to the $\pi\pi$ scattering cross section~\cite{Donoghue:1990xh,Gasser:1990bv,Colangelo:2001df}. Furthermore, it enters the chiral expansion of
the pion decay constant in two-flavor chiral perturbation theory \cite{Gasser:1983kx}
\begin{equation}1
 \frac{F_\pi}{\chiral{F}_\pi} = 1 + \frac{1}{6}M_\pi^2 \langle r_S^2 \rangle_\pi^l + \frac{13 M_\pi^2}{192\pi^2 F_\pi^2} + \mathcal{O}(M_\pi^4)  \,.
 \label{eq:F_pi_chiral_expansion}
\end{equation}
The next-to-leading order expression for $\langle r^2_S \rangle_\pi^l$ depends only on a single low-energy constant $\bar{l}_4$,
\begin{equation}
 \langle r_S^2 \rangle_\pi^l = \frac{1}{(4\pi f)^2} \l[-\frac{13}{6} + \l(6\bar{l}_4 + \log\l(\frac{M_{\pi,\mathrm{phys}}^2}{M_\pi^2}\r)\r) \r]\,.
  \label{eq:rsqr_chiral_expansion}
\end{equation}
In principle, this allows to obtain a rather straightforward, model-independent determination of $\bar{l}_4$ from lattice QCD. Nevertheless, only few modern lattice calculations exist in literature
\cite{Gulpers:2013uca,Gulpers:2015bba,Koponen:2015tkr}, arguably none of them with fully controlled systematics. A complication arises from the notorious quark-disconnected contributions to the pion scalar form factor which render such a calculation
computationally and conceptually more demanding than e.g. a study of vector form factors. \par

Beyond the two-flavor theory contributions of the strange quark to the pion scalar form factor are conveniently parameterized in terms of $\SU{3}_F$ octet and singlet combinations. Lattice results for the
corresponding radii $\langle r_S^2 \rangle_\pi^8$ and $\langle r_S^2 \rangle_\pi^0$ are even more scarce \cite{Koponen:2015tkr} than for $\langle r_S^2 \rangle_\pi^l$ as effects of the strange quark are 
mediated entirely by quark-disconnected diagrams. Still, due to the relatively large quark-disconnected contributions particularly at zero and small momentum transfer, the effects of the strange quark 
are not negligible, leading to an expected hierarchy of the radii, i.e. $\langle r_S^2 \rangle_\pi^8 < \langle r_S^2 \rangle_\pi^l < \langle r_S^2 \rangle_\pi^0$ assuming  $m_l<m_s$. \par


\section{Lattice setup}
Our lattice calculations are carried out on gauge ensembles with $N_f=2+1$ flavors of non-perturbatively $\mathcal{O}(a)$-improved Wilson fermions~\cite{Sheikholeslami:1985ij} and the tree-level 
Symanzik-improved L\"uscher-Weisz gauge action~\cite{Luscher:1984xn}. These ensembles have been generated by the Coordinated Lattice Simulation (CLS) consortium~\cite{Bruno:2014jqa} employing a twisted mass
regulator in the (mass-degenerate) light quark sector to suppress exceptional configurations~\cite{Luscher:2012av} and the rational approximation~\cite{Clark:2006fx} for the dynamic strange quark. For the 
required reweighting of physical observables we employ reweighting factors computed using exact low mode deflation, cf. Ref.~\cite{Kuberski:2023zky}. The only exception is the E300 ensemble, for which 
stochastic estimators have been used as discussed in Ref.~\cite{Bruno:2014jqa}. An overview of the 13 ensembles currently used in this study can be found in Table~\ref{tab:ensembles}. For further details we 
refer to Ref.~\cite{Bruno:2014jqa}. Moreover, we remark that we make use of the procedure developed in Ref.~\cite{Mohler:2020txx} to deal with positivity violations of the fermion determinant that affect a 
small subset of gauge configurations on some ensembles. At present we have only included ensembles that lie on the chiral trajectory defined by the constraint $\tr[M] = 2m_l+m_s =\mathrm{const}$ for the 
quark mass matrix $M$. \par

Whenever conversion to physical units is required, we use the values for $t_0^\mathrm{sym}/a^2$ from Ref.~\cite{Bruno:2016plf} determined at the symmetric point together with the world average for 
the gradient flow scale $t_0$ \cite{Luscher:2010iy} given by FLAG in Ref.~\cite{FlavourLatticeAveragingGroupFLAG:2021npn}, i.e. $t_0^\phys=0.14464(87)$. At this stage of the analysis we do not yet 
carry out the physical extrapolation of our lattice results, hence the scale only enters in the values of $M_\pi$ in Table~\ref{tab:ensembles} and in results for scalar radii in physical units on individual 
ensembles. Besides, the values for the lattice spacings quoted in Table~\ref{tab:ensembles} corresponding to the four values of $\beta\in\l\{3.40,3.46,3.55,3.70\r\}$ have been computed from this procedure 
as well. \par

\begin{table}[!t]
 \caption{List of CLS gauge ensembles used in this work. Open and periodic boundary conditions in time are indicated by superscripts ``$o$'' and ``$p$'', respectively. $N_\mathrm{conf}$ is the number of
 gauge configurations on which the $N_\mathrm{meas}^\mathrm{3pt}$ ($N_\mathrm{meas}^\mathrm{2pt}$) measurements of three-point (two-point) functions have been performed. $\Delta N_\mathrm{conf}$ denotes
 the index stride between gauge configurations with measurements and the last column lists the available values of $\tsep/a$. } 
 \setlength{\tabcolsep}{0.28em}
 \centering
 \begin{tabular*}{\textwidth}{ccrrcrcrrcc}
  \hline\hline
  ID$^{BC}$ & $a$/fm& T/a & L/a& $M_\pi / \mev$ &  $N_\mathrm{conf}$ & $\Delta N_\mathrm{conf}$ & $N_\mathrm{meas}^\mathrm{3pt}$ & $N_\mathrm{meas}^\mathrm{2pt}$ & $\tsep/a$ \\
  \hline\hline                                                            
  C101$^o$  & 0.086 &  96 & 48 & 222 & 1973 & 1 & 15784 & 276220 & 11, 15, 19, 23, 27, 31, 35 \\     
  N101$^o$  &       & 128 & 48 & 279 & 1260 & 1 & 10080 & 176400 &  \\                               
  H105$^o$  &       &  96 & 32 & 285 & 1011 & 1 &  8088 & 141540 &  \\                               
  H102$^o$  &       &  96 & 32 & 351 & 1020 & 1 &  8160 & 142800 &  \\                               
  \hline
  D450$^p$  & 0.076 & 128 & 64 & 218 &  497 & 1 &  3976 &  63616 & 13, 17, 21, 25, 29, 33, 37, 41 \\ 
  N451$^p$  &       & 128 & 48 & 287 & 1003 & 1 &  8024 & 128384 &  \\                               
  S400$^o$  &       & 128 & 32 & 357 &  994 & 2 &  7952 & 127232 &  \\                               
  \hline
  E250$^p$  & 0.064 & 192 & 96 & 132 &  491 & 2 &  3928 &  62848 & 15, 21, 27, 33, 39, 45, 51 \\     
  D200$^o$  &       & 128 & 64 & 202 &  991 & 2 &  7928 & 138740 &  \\                               
  N200$^o$  &       & 128 & 48 & 285 & 1705 & 1 & 13640 & 238700 &  \\                               
  N203$^o$  &       & 128 & 48 & 346 & 1535 & 1 & 12280 & 214900 &  \\                               
  \hline
  E300$^o$  & 0.049 & 192 & 96 & 174 &  569 & 2 &  4552 &  79660 & 19, 27, 35, 43, 51, 59, 67 \\     
  J303$^o$  &       & 192 & 64 & 256 & 1068 & 1 &  8544 & 149520 &  \\                               
  \hline\hline
 \end{tabular*}
 \label{tab:ensembles}
\end{table}

\subsection{Computational details}
\begin{figure}[thb]
 \centering
 \includegraphics[width=.275\textwidth]{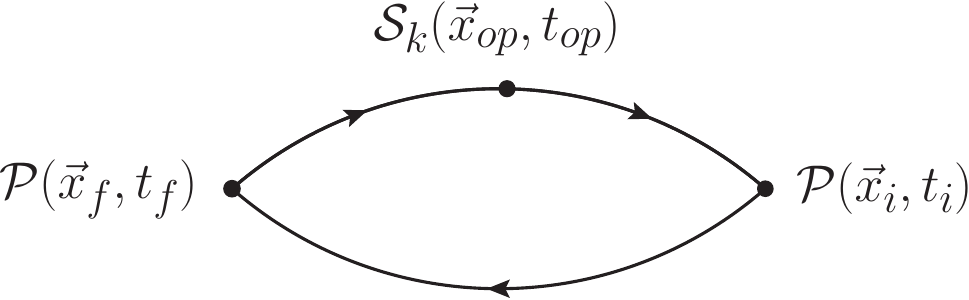} \bf $\begin{array}{c} \vspace{1em} + \\ \\ \end{array}$ 
 \includegraphics[width=.275\textwidth]{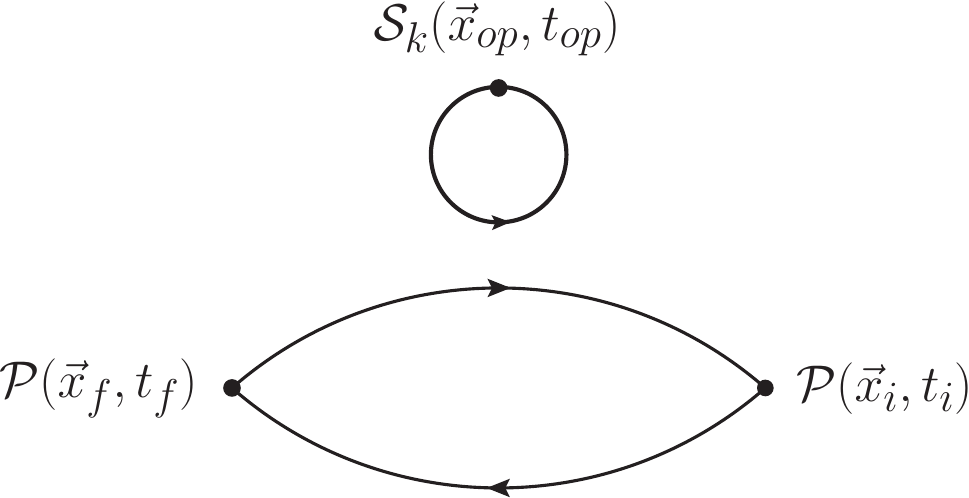} \bf $\begin{array}{c} \vspace{1em} - \\ \\ \end{array}$ 
 \includegraphics[width=.275\textwidth]{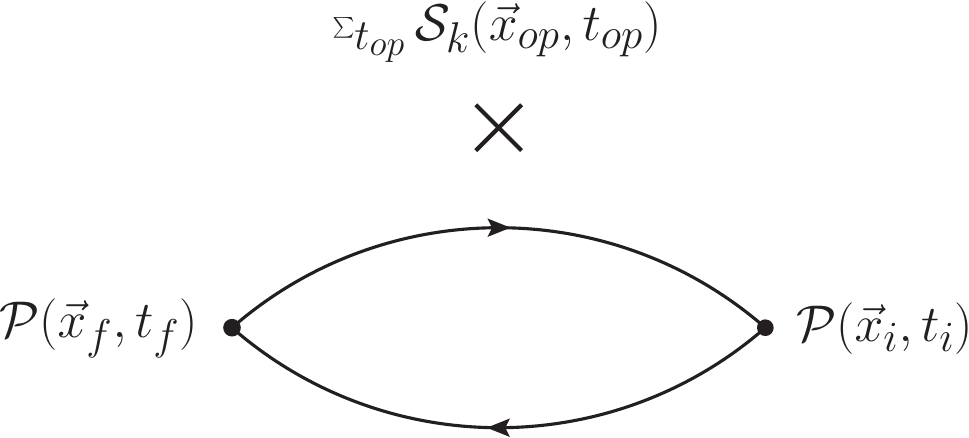} \\
 \vspace{-1.5em}
 \caption{Quark-connected and disconnected contributions to the three-point function in Eq.~(\ref{eq:3pt}). The last diagram contributes only for vanishing loop momentum.}
 \label{fig:3pt}
\end{figure}

A lattice study of the pion scalar form factor requires the evaluation of quark-connected and disconnected three-point functions depicted in Fig.~\ref{fig:3pt}.
The basic building blocks are quark-disconnected loops 
\begin{equation}
 C^\mathrm{1pt}(\vp, t) = \sum_{\mathbf{x}} e^{i\vp\cdot\mathbf{x}} \langle\mathcal{S}_f(\mathbf{x}, t)\rangle_F \,, \notag
 \label{eq:1pt}
\end{equation}
and quark-connected two- and three-point functions
\begin{align}
 C^\mathrm{2pt}(\vp, \vxi, t_f-t_i) &= \sum_{\vxf} e^{i \vp\cdot(\vxf - \vxi)} \langle \mathcal{P}(\vxf-\vxi, t_f-t_i) \mathcal{P}^\dag(\mathbf{0}, 0) \rangle \,,
 \label{eq:2pt} \\
 C^\mathrm{3pt}_k(\vpf, \vq, \vxi, \tsep, \tins ) &= \sum_{\vxf,\vxop} e^{i \vpf \cdot (\vxf-\vxi)} e^{i\vq \cdot (\vxop-\vxi)} \langle \mathcal{P}(\vxf-\vxi,
 \tsep) \mathcal{S}_k(\vxop-\vxi, \tins) \mathcal{P}^\dag(\mathbf{0}, 0) \rangle\,, \label{eq:3pt}
\end{align}
where $\mathcal{P}(x)=u(x)\g{5}\bar{d}(x)$ denotes the interpolating operator for the pion and $\tsep=t_f-t_i$, $\tins=t_{op}-t_i$ are shorthands for the pertinent Euclidean time-separations. In
addition to a scalar operator insertion involving only light quarks, $\mathcal{S}_l = \bar{u}u+\bar{d}{d}$, we also compute the octet and singlet flavor combinations involving the strange quark, 
i.e. $\mathcal{S}_8 = \bar{u}u+\bar{d}{d}-2\bar{s}s$ and $\mathcal{S}_0 = \bar{u}u+\bar{d}{d}+\bar{s}s$. \par

For the computation of the quark-disconnected loops we employ stochastic volume sources and the one-end~trick~\cite{Jansen:2008wv} to obtain a frequency splitting estimator~\cite{Giusti:2019kff} in 
combination with the generalized hopping parameter expansion \cite{Gulpers:2013uca} and hierarchical probing \cite{Stathopoulos:2013aci}. For further details on the setup we refer to Ref.~\cite{Ce:2022eix}. \par

The quark-connected two- and three-point functions are evaluated on and averaged over point sources. The actual source setup depends on the boundary conditions used in time on a given ensemble, cf.
Table~\ref{tab:ensembles}. For ensembles with periodic boundary conditions (pBC) sources are drawn randomly without replacement with the only constraint arising from the combination of the truncated solver
method with the Schwartz alternating procedure (SAP) preconditioning \cite{Luscher:2003vf,vonHippel:2016wid}. The three-point functions are computed using the sequential sink method with the final state 
either produced at rest or with one unit of momentum $\vpf=(1,0,0)^T$. For ensembles with pBC we use eight sources per gauge configuration for the three-point functions, which is reflected by the numbers for
$N_\mathrm{meas}^\mathrm{3pt}$ in Table~\ref{tab:ensembles}. The forward propagators are re-used for all values of $\tsep$ as well as for the computation of two-point functions. Additional 120 sources 
per gauge configuration are used to increase the two-point function statistics for the computation of quark-disconnected (2+1)-point functions, resulting in much larger numbers for 
$N_\mathrm{meas}^\mathrm{2pt}$. For ensembles with open boundary conditions (oBC) four sources are randomly placed on each of the two possible source timeslices symmetric around $T/2$ for each value of 
$\tsep$ for the computation of three-point functions. Additional two-point functions are computed on the same set of timeslices to achieve a similar $N_\mathrm{meas}^\mathrm{2pt}/N_\mathrm{conf}$ ratio
as for ensembles with pBC. \par

\begin{figure}[thb]
 \centering
 \includegraphics[totalheight=0.20\textheight]{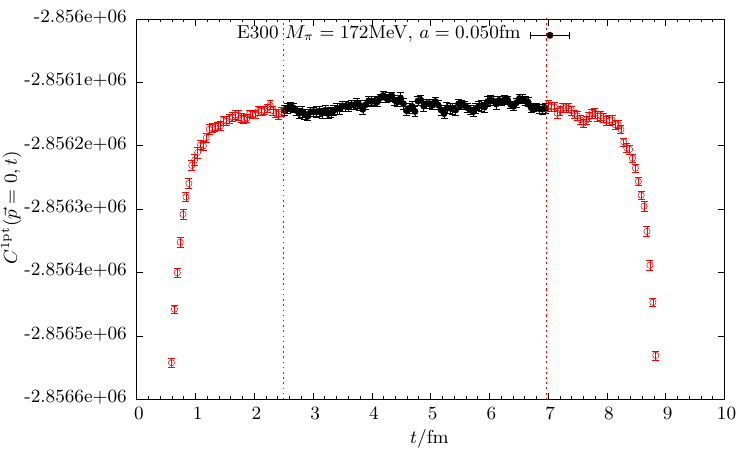}
 \includegraphics[totalheight=0.20\textheight]{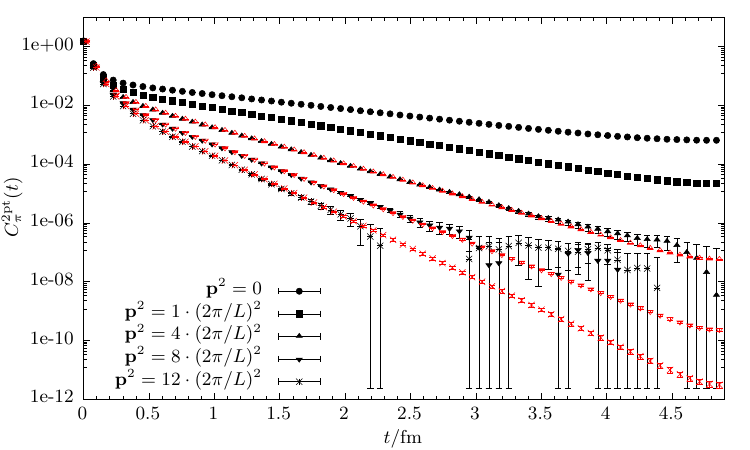}
 \caption{Left panel: Scalar one-point function on the E300 ensemble with open boundary conditions in time. Red data are excluded in the construction of quark-disconnected (2+1)-point functions. Right
 panel: Pseudoscalar two-point functions for several momenta on the D450 ensemble. For $\vp^2 \geq 4\cdot(2\pi/L)^2$ the large-$t$ tail has been reconstructed (red points) from a fit to data at smaller
 values of $t$. Corresponding data points are horizontally displaced for legibility.}
 \label{fig:E300_1pt_D450_2pt}
\end{figure}

However, a subtlety arises for the computation of quark-disconnected (2+1)-point functions on ensembles with oBC. While for pBC two-point functions computed on any point source can be used in forward 
and backward time direction, such averaging is limited on ensembles with oBC as a safe distance from the boundaries needs to be kept for the operator insertion in $t_{op}$ to avoid contamination by 
boundary effects. The range that needs to be excluded can be readily assessed from the time-dependence of the scalar loop at zero-momentum transfer and can be as large as $t_\mathrm{ex}=2.5\fm$ depending 
on $M_\pi$ as shown in the left panel of Fig.~\ref{fig:E300_1pt_D450_2pt} for the most chiral ensemble with oBC. In particular, at larger values of $\tsep$ and for ensembles with smaller physical time 
extent this can lead to greatly reduced effective statistics for these measurement compared to ensembles with pBC. \par

Statistical errors are computed using the binned jackknife method, and we adopt the procedure previously used in Refs.~\cite{Agadjanov:2023jha,Djukanovic:2023beb,Djukanovic:2023jag} for the computation of
isoscalar observables in the nucleon sector to identify individual measurements on a few point sources that represent extreme outliers and would cause unduly inflated statistical errors. The 
affected gauge configurations have been excluded from further analysis which is taken into account in the numbers for $N_\mathrm{conf}$ in Table~\ref{tab:ensembles}. \par

\section{Form factor extraction} \label{sec:FF_extraction}
\begin{figure}[thb]
 \centering
 \includegraphics[totalheight=0.20\textheight]{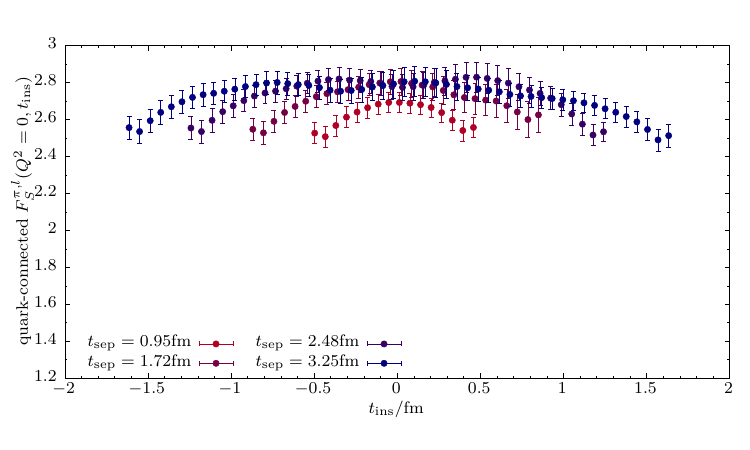}
 \includegraphics[totalheight=0.20\textheight]{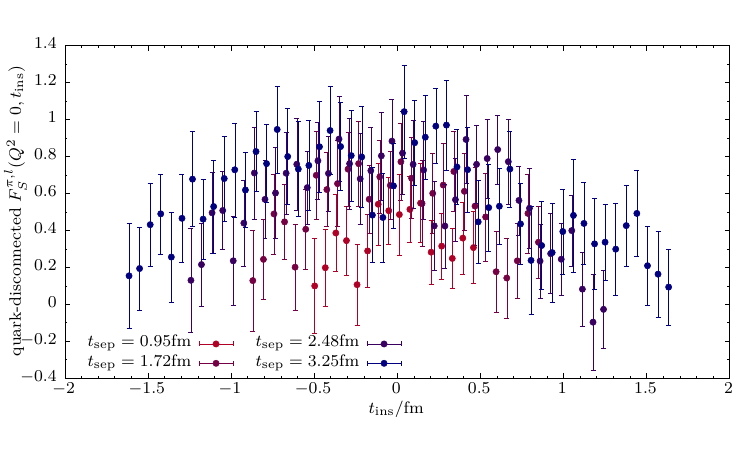} \\
 \includegraphics[totalheight=0.20\textheight]{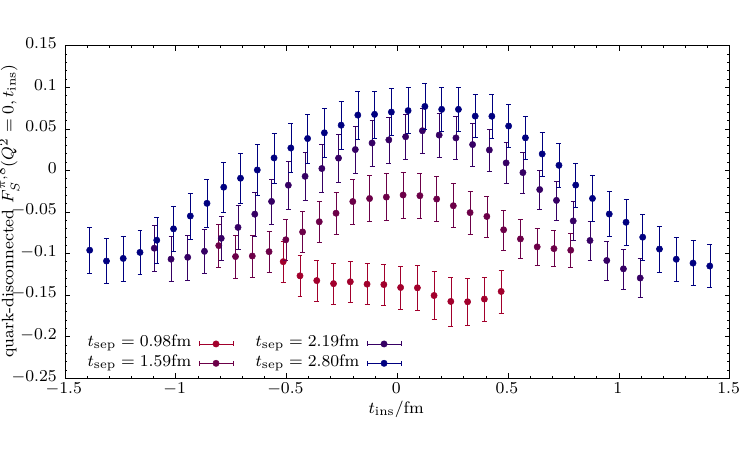}
 \includegraphics[totalheight=0.20\textheight]{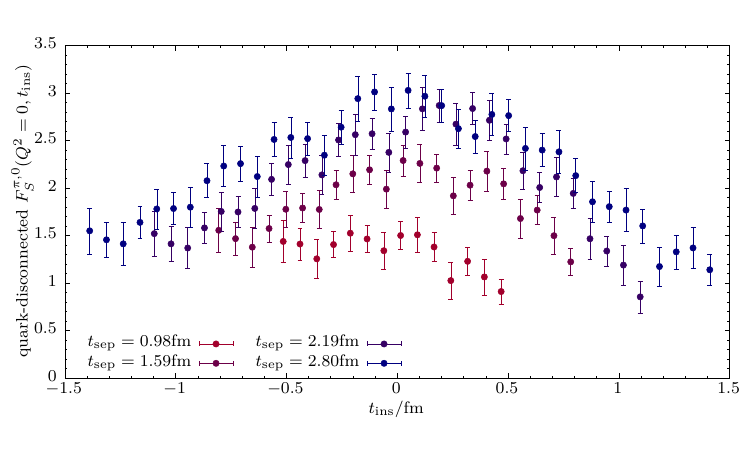}
 \caption{Effective form factors at $Q^2=0$ with $\vpf=(0,0,0)^T$. Upper row: Light quark-connected (left panel) and quark-disconnected contribution on the E250 ensemble. Lower row: Octet (left panel) and
 singlet (right panel) quark-disconnected contribution on the N451 ensemble. Data for every second available value of $\tsep$ are included.}
 \label{fig:eff_FF}
\end{figure}

In a first step to extract the scalar matrix element we compute effective form factors from the standard ratio of two- and three-point functions
\begin{align} 
 R(p_f^2,q^2,p_i^2,& t_f-t_i, t_{op}-t_i) =  \notag \\
  & \frac{C_3(p_f^2,q^2,p_i^2, t_f-t_i, t_{op}-t_i)}{C_2(p_f^2,t_f-t_i)} \cdot \sqrt{\frac{C_2(p_i^2,t_f-t_{op}) C_2(p_f^2, t_{op}-t_i) C_2(p_f^2, t_f-t_i) }{ C_2(p_f^2,t_f-t_{op}) C_2(p_i^2, t_{op}-t_i) C_2(p_i^2, t_f-t_i)}} \,.
 \label{eq:ratio}
\end{align}
which yields the desired ground state matrix elements $\langle\pi(p_f^2) |\mathcal{S}_k(q^2)| \pi(p_i^2)\rangle \sim F_S^{\pi,k}(Q^2)$ at asymptotically large Euclidean time separations. While the two- 
and three-point functions for the pion generally exhibit a constant signal-to-noise ratio at zero-momentum, they develop a signal-to-noise problem at increasing momentum transfer, as shown in the right
panel of Fig.~\ref{fig:E300_1pt_D450_2pt} for the the case of the two-point functions on D450 . In order to alleviate this issue in the construction of the ratio in Eq.~(\ref{eq:ratio}) for larger values
of $\tsep$ we fit the ground-state contribution to the two-point functions at $p^2\ge 2\cdot (2\pi/L)^2$ and replace the corresponding two-point function data by the fit result, which is illustrated by 
the red data points in the right panel of Fig.~\ref{fig:E300_1pt_D450_2pt}. Example data for the light, quark-connected contribution as well as the quark-disconnected light, octet and singlet contributions 
to the effective form factors at vanishing transfer and $\vpf=(0,0,0)^T$ are shown in Fig.~\ref{fig:eff_FF}. The quark-disconnected contributions for $F^{\pi,l}_S$ and $F^{\pi,0}_S$ can reach similar size 
as the quark-connected piece at $q^2=0$ depending on the lattice spacing and light quark mass, in agreement with expectations from chiral perturbation theory~\cite{Juttner:2011ur}. However, their relative 
(and absolute) size decreases at increasing momentum transfer. \par 

In order to gain further suppression of excited states we employ the summation method~\cite{Maiani:1987by,Dong:1997xr,Capitani:2012gj}
\begin{align}
 S(p_f^2,q^2,p_i^2, \tsep) &= \sum_{\tins=t_0}^{\tsep-t_0} R(p_f^2,q^2,p_i^2, \tsep, \tins) \notag \\
  &= \mathrm{const} + \langle\pi(p_f^2) | S(q^2)| \pi(p_i^2)\rangle (\tsep-t_0) + \mathcal{O}(e^{-\Delta\tsep}) \,,
 \label{eq:summation_method}
\end{align}
where $\Delta$ is the mass gap between ground and first excited state and we choose $t_0=0.4\fm$.  We find this choice to improve the signal quality for non-vanishing source and sink momenta, while still
leaving enough points in the sum at smaller values of $\tsep$. For the preliminary analysis in this proceedings contribution we include data for $\tsep\gtrsim 1.0\fm$ up to $\tsep\lesssim 3.5\fm$. However,
the application of the summation method turns out to be challenging on a few ensembles with oBC due to the aforementioned, drastic reduction in effective statistics at $\tsep\gtrsim 2\fm$ in the
disconnected contribution depending on the excluded time range $t_\mathrm{ex}$ next to the boundary and the value of $T$. On the affected ensembles we observe a systematic deviation at small $Q^2$ for
$\tsep\gtrsim 2\fm$ from the expected linear behavior in Eq.~(\ref{eq:summation_method}) towards larger slopes, which leads to an overestimation of $F_S^{\pi,k}(Q^2)$. Since the quark-connected contribution 
is not subjected to such a reduction in statistics as function of $\tsep$, the effect is most prominent in $F_S^{\pi,l}(Q^2)$ and $F_S^{\pi,0}(Q^2)$, whereas the octet combination $F_S^{\pi,8}(Q^2)$ 
is less affected. In the future we intend to investigate alternative ways to achieve ground-state domination, e.g. simultaneous multi-state fits directly to the two- and three point functions, as well as 
optimizing the choice of the values of $\tsep$ used in the analysis. Besides, the number of two-point function measurements on the affected ensembles may be further increased to compensate for this 
effect. \par

\begin{figure}[thb]
 \centering
 \includegraphics[totalheight=0.20\textheight]{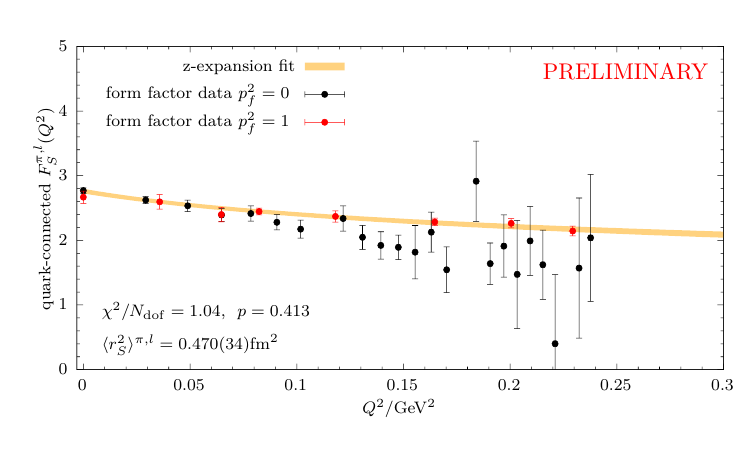}
 \includegraphics[totalheight=0.20\textheight]{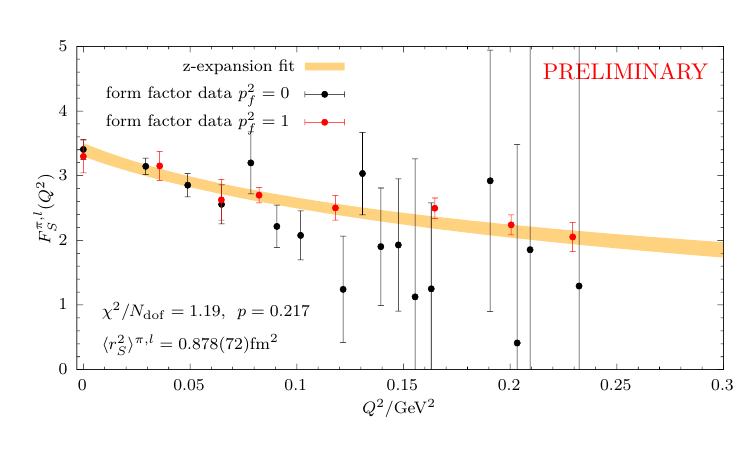} \\
 \includegraphics[totalheight=0.20\textheight]{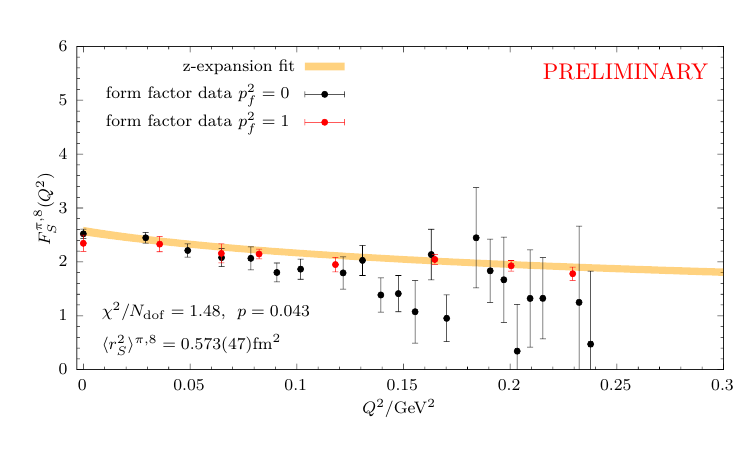}
 \includegraphics[totalheight=0.20\textheight]{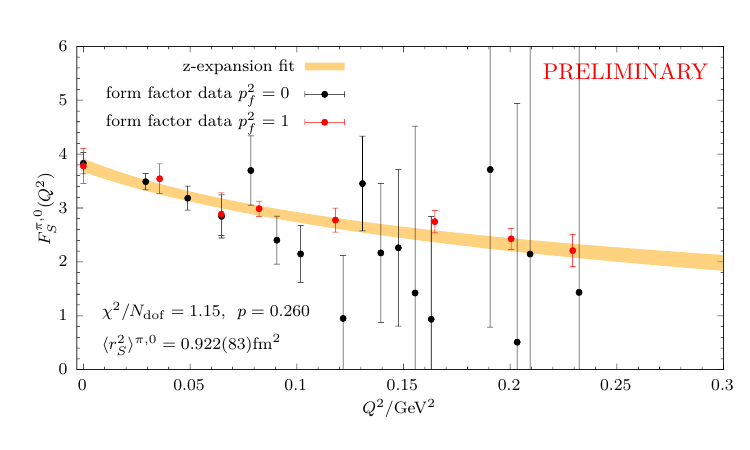}
 \caption{Unrenormalized form factor data from summation method (see text) and first order $z$-expansion fits on the E250 ensemble at physical light quark mass. Results are shown for the quark-connected 
 and full light quark contribution (upper two panels), as well as the octet and singlet combinations (lower two panels). Only data for $\vpi^2<4\dot (2\pi/l)^2$ are shown in the plots, as the signal 
 quality rapidly deteriorates for increasing source momenta.}
 \label{fig:E250_FF_z_expansion}
\end{figure}

\section{Scalar radius}
Previous lattice studies had only access to very few momenta~\cite{Gulpers:2013uca,Gulpers:2015bba,Koponen:2015tkr}, hence they essentially used the slope between the form factor at $Q^2=0$ and the first 
one or two non-vanishing momenta to estimate the corresponding radii. The situation is quite different in the present study, as our setup allows us to compute the pion scalar form factor for a large number 
of different momenta with unprecedented momentum resolution on the ensembles with largest physical volume, as can be seen in Fig.~\ref{fig:E250_FF_z_expansion} for the E250 ensemble at physical quark mass. 
The three-point function measurements with one unit of sink momentum yield very precise results for a large range of momenta, while further refining the resolution at small momentum transfer, which is 
particularly beneficial for the computation of radii. In order to parametrize the form factor data in model-independent way we employ the $z$-expansion~\cite{Lee:2015jqa}
\begin{equation}
 F_S^{\pi,k}(Q^2) = \sum_{n=0}^{N_z} a_n z^n\,, \qquad z=\frac{\sqrt{t_\mathrm{cut}+Q^2} - \sqrt{\vphantom{Q^2}t_\mathrm{cut}-t_0}}{\sqrt{t_\mathrm{cut}+Q^2} + \sqrt{\vphantom{Q^2}t_\mathrm{cut}-t_0}} \, 
 \label{eq:z_expansion}
\end{equation}
where we have $t_\mathrm{cut}=4M_\pi^2$ and use the ``optimal'' choice for $t_0=t_\mathrm{cut}(1-\sqrt{1+Q^2_\mathrm{max}/t_\mathrm{cut}})$. At the present level of statistical precision we find that
$N_Z=1$ yields a good description of the form factor data up to $Q^2\leq 0.3\gev^2$. The scalar radii $\langle r_S^2\rangle_\pi^k$ are obtained from the $a_1$ coefficient in Eq.~(\ref{eq:z_expansion}).
Examples for the corresponding fits on our most chiral ensemble are included in Fig.~\ref{eq:z_expansion} together with the results for the radii. The expected hierarchy of radii, i.e. $\langle r_S^2
\rangle_\pi^8 < \langle r_S^2 \rangle_\pi^l < \langle r_S^2 \rangle_\pi^0$ is reproduced by our lattice data. There is no clear indication of lattice artifacts or finite volume effects in the data. However, 
obtaining final physical results for radii and $\bar{l}_4$ with fully controlled systematics will require a more detailed analysis of excited states and the $Q^2$-dependence of the form factor data, as well 
as further increasing statistics on several ensembles.  \par

\begin{figure}[thb]
 \centering
 \includegraphics[totalheight=0.20\textheight]{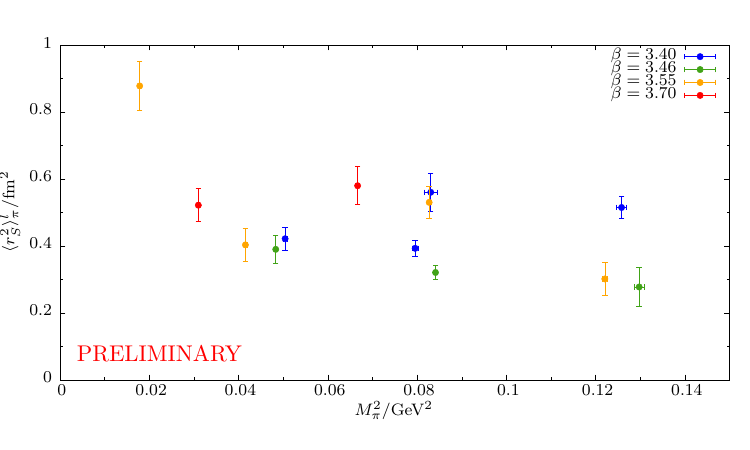}
 \includegraphics[totalheight=0.20\textheight]{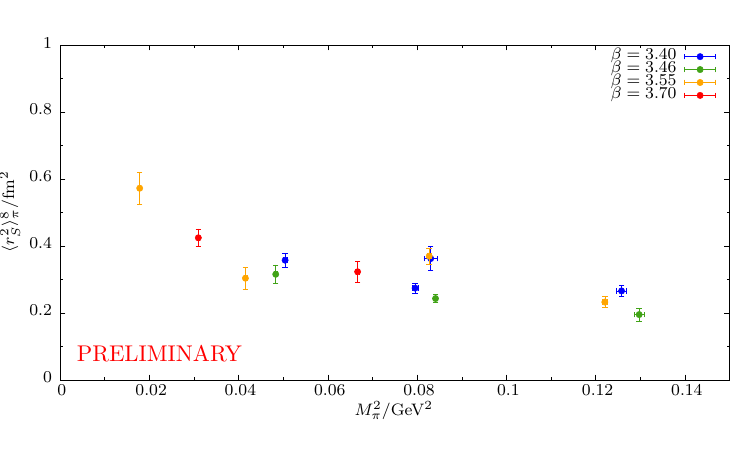} \\
 \includegraphics[totalheight=0.20\textheight]{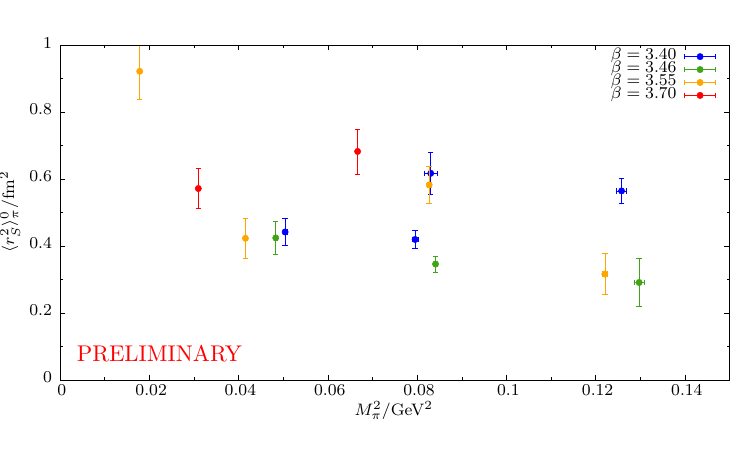}
 \caption{Results for $\langle r_S^2 \rangle_\pi^l$, $\langle r_S^2 \rangle_\pi^8$ and $\langle r_S^2 \rangle_\pi^0$ in physical units as a function of $M_\pi^2$.}
 \label{fig:chiral_behavior}
\end{figure}

Fig.~\ref{fig:chiral_behavior} shows the chiral behavior of the lattice data for $\langle r_S^2\rangle_\pi^l$, $\langle r_S^2 \rangle_\pi^8$ and $\langle r_S^2 \rangle_\pi^0$. The data for 
$\langle r_S^2\rangle_\pi^l$ are compatible with the expected logarithmic behavior for in Eq.~(\ref{eq:rsqr_chiral_expansion}). However, a few ensembles at heavier pion mass deviate from the expected 
curve for $\langle r_S^2 \rangle_\pi^l$ and $\langle r_S^2 \rangle_\pi^0$, which is caused by still insufficient effective statistics for the (2+1)-point functions at larger values of $\tsep$ on these 
ensembles, as discussed in section~\ref{sec:FF_extraction}. For the octet combination the effect is much less significant. \par

\section{Outlook}
In this proceedings contribution we have presented the status of our ongoing lattice study of the pion scalar form factor on a set of currently 13 CLS gauge ensembles. While the data production is not yet 
complete, and the physical extrapolation including a full error budget is still work in progress, we find that results on individual ensembles are already very competitive in terms of statistical errors
when compared to previous lattice calculations. Furthermore, we achieve an unprecedented momentum resolution on boxes with large physical volumes near the physical point, which should in principle allow us 
to achieve a much more comprehensive assessment of systematic effects in the computation of radii. \par
In the future, we intend to double the number measurements for two- and three-point functions on our three most chiral ensembles (i.e. E250, E300 and D200) for which measurements are currently only
available on every second gauge configuration. In addition, the number of two-point function measurements for the computation of quark-disconnected (2+1)-point functions may be further increased on selected 
ensembles. This concerns particularly ensembles with open boundary conditions and smaller values of $T$, for which statistics for these contributions is currently quite limited at larger values of $\tsep$. 
Moreover, we plan to extend our study to CLS ensembles on a different chiral trajectory with $m_s=\mathrm{phys}$ to gain even better control over quark mass dependence for 
$\langle r_S^2 \rangle_\pi^8$ and $\langle r_S^2 \rangle_\pi^0$ when performing physical extrapolations. \par

\section*{Acknowledgments}
This resaerch is supported by the Deutsche Forschungsgemeinschaft (DFG, German Research Foundation) through project HI~2048/1-3 (project No.~399400745). The authors gratefully acknowledge the Gauss Centre for Supercomputing e.V. (www.gauss-centre.eu) for funding this project by providing computing time on the GCS Supercomputer SuperMUC-NG at Leibniz Supercomputing Centre and on the GCS Supercomputers JUQUEEN\cite{juqueen} and JUWELS\cite{JUWELS} at Jülich Supercomputing Centre (JSC). Additional calculations have been performed on the HPC clusters Clover at the Helmholtz-Institut Mainz and Mogon II and HIMster-2 at Johannes-Gutenberg Universit\"at Mainz. We thank our colleagues in the CLS initiative for sharing gauge ensembles.

\bibliographystyle{JHEP.bst}

\providecommand{\href}[2]{#2}\begingroup\raggedright\endgroup
\end{document}